\documentclass[fleqn,12pt,twoside]{article}
\usepackage{espcrc1,epsfig}

\usepackage{graphicx}
\usepackage[figuresright]{rotating}

\newcommand{\beq} {\begin{equation}}
\newcommand{\eeq} {\end{equation}}
\newcommand{\beqa} {\begin{eqnarray}}
\newcommand{\eeqa} {\end{eqnarray}}
\newcommand{\mrm}[1] {{\mathrm{#1}}}
\newcommand{\ie}{{\it i.e.}}
\newcommand{\eg}{{\it e.g.}}
\newcommand{\cf}{{\it cf.}}
\newcommand{\etal}{{\it et al.}}
\newcommand{\gev}{{\mrm{GeV}}}
\newcommand{\as}{\alpha_s}

\newcommand{\ieps}{i\varepsilon}

\newcommand{\morder}[1]{{\cal O}\left(#1 \right)}
\newcommand{\eq}[1]{(\ref{#1})}

\newcommand{\ave}[1]{\langle{#1}\rangle}

\newcommand{\qu}{{\rm q}}
\newcommand{\qb}{{\rm\bar q}}
\newcommand{\qq}{\qu\qb\ }

\newcommand{\gsim}{\buildrel > \over {_\sim}}
\newcommand{\lsim}{\buildrel < \over {_\sim}}
\newcommand{\ol}{\overline}     

\newcommand{\Slash}[1]{ \parbox[b]{0.6em}{$#1$} \hspace{-0.55em}
                            \parbox[b]{0.55em}{ \raisebox{-0.2ex}{$/$}}}

\newcommand{\PL}[3]{Phys.~Lett.~{{#1}}~({#3})~{#2}}
\newcommand{\NP}[3]{Nucl.~Phys.~{{#1}}~({#3})~{#2}}
\newcommand{\PRD}[3]{Phys.~Rev.~{D{#1}}~({#3})~{#2}}
\newcommand{\PRL}[3]{Phys.~Rev.~Lett.~{{#1}}~({#3})~{#2}}

\newcommand{\PRe}[3]{Phys.~Rep.~{{#1}}~({#3})~{#2}}

\newcommand{\AmS}{{\protect\the\textfont2
  A\kern-.1667em\lower.5ex\hbox{M}\kern-.125emS}}

\title{Probing QCD at Medium $Q^2$\footnote{Introductory talk at the ``European Workshop on the QCD Structure of the Nucleon'' in Ferrara, Italy (April 2002). Transparencies including figures may be found at http://www.fe.infn.it/qcd-n02/talks.htm}}

\author{Paul Hoyer\thanks{On leave from the Department of Physics,
University of Helsinki, Finland. Adjoint senior scientist, Helsinki Institute of Physics. Research supported in part by the
European Commission under contract HPRN-CT-2000-00130.}
\address{Nordita, Blegdamsvej 17, DK-2100 Copenhagen, Denmark\\
	E-mail: {\tt hoyer@nordita.dk}}}
       
\begin{document}

% typeset front matter
\maketitle

\begin{abstract} Quantum Chromodynamics (QCD) is a firmly established part of the Standard Model, yet its long distance properties remain challenging at a conceptual level. In recent years significant experimental and theoretical progress has been made towards a field theoretic understanding of hadrons as QCD bound states, addressing the complex properties of relativistic bound states in a gluon condensate with spontaneously broken chiral symmetry. In this introductory talk I give a personal view of some of the issues involved: Measuring hadron matrix elements, understanding their QCD dynamics and explaining the apparently close relationship between hard and soft processes.
\end{abstract}

\vspace{1.5cm}
{\bf CONCEPTUAL ISSUES}
\vspace{.5cm}

A principal goal of hadron physics is the field theoretic understanding of QCD bound states. Compared to QED bound states (positronium) we face a number of highly non-trivial issues:
\begin{itemize}

\item {\em Relativistic motion.} Most of the proton mass is dynamic, with the current quark masses contributing just a few MeV. The relativistic nature of hadrons is reflected in the broad momentum distribution of the quarks, and in the gluons carrying half the total momentum. Our understanding of positronia is based on the perturbative expansion and relies on the smallness of the mean velocity $v/c = \alpha \simeq 1/137$ of the constituents.

\item {\em The gluon condensate.} Unlike the situation in QED, the QCD vacuum is complex and contains a `condensate' of gluons. Hadrons are immersed in this condensate and strongly influenced by it. There are interference effects between soft gluons in the condensate and in the hadron wave function. It is likely that the condensate is at the origin of color confinement and the absence of free quark states.

\item {\em Chiral symmetry breaking.} The QCD vacuum breaks the symmetry between (massless) left- and right-handed quarks. Consequently the pion (Goldstone boson) mass arises solely from the explicit breaking of chiral symmetry due to the finite current quark masses. There is no analogy of this in QED, where the scale is set by the electron mass.
 
\end{itemize}

It is thus clear that a QCD analysis of hadrons requires concepts and methods that go beyond those with which we are familiar in QED bound state studies and in perturbative QCD (PQCD). In contrast to the generation of quark and lepton masses via the Higgs mechanism, our understanding of the proton mass requires conceptual advances. This, and the fact that QCD dynamics is responsible for $\gsim 98\%$ of the proton mass, is what makes the study of hadron structure so fascinating.

In our ambitious quest for hadron structure we are fortunate to know the basic quark and gluon interactions from ${\cal L}_{QCD}$. In particular, this allows numerical studies using the methods of lattice QCD. Furthermore, moderate momentum transfers $Q^2 \lsim 10\ \gev^2$ are sufficient to explore hadron structure at a resolution $\sim 0.1$ fm. Hence a broad range of measurements are feasible at present and future facilities. These aspects make the analysis of hadron structure practical and more systematic than physics beyond the standard model, where the interactions are unknown and experiments difficult.

An understanding of hadron structure does not imply a full `solution' of QCD. Comparison between theory and experiment was required to establish that perturbative QED methods are adequate for calculating positronium properties with high precision. In effect, data selected a useful approximation scheme, which would have been difficult to defend on purely theoretical grounds. It is not  {\it a priori} clear at what precision the (asymptotic) QED perturbative series breaks down, nor for what range of $\alpha$ it is at all adequate. In QCD we may analogously search for an expansion that is formally exact and provides a reasonable approximation of measurements at low orders.

Data already provides many hints. The success of the constituent quark model makes it a good candidate for a lowest order approximation of QCD bound states. The similarity of the hadron spectrum to that of QED suggests the use of PQCD. Early scaling and the success of simple phenomenological extrapolations from high to low $Q^2$ also suggests a close relation between the short distance (perturbative) and long distance (perhaps also perturbative?) regimes. The insight that the strong coupling may freeze at $\as(Q^2=0) \simeq 0.5$, thus making PQCD relevant at all $Q^2$, has been coined the `QCD Faith Transition' \cite{yd}.

My presentation is divided into three `steps', representing various levels in the study of hadron structure.

\section{First step: MEASURING MATRIX ELEMENTS}

\subsection{Inclusive Deep Inelastic Scattering}

Data on hadron structure comes primarily from high energy collisions. In order to separate the collision from the bound state dynamics we use QCD factorization theorems \cite{css}. The classic example is deep inelastic lepton scattering (DIS), $ep\to e+X$, in the Bjorken limit of high photon virtuality $Q^2$ and energy $\nu$ in the target proton rest frame. The DIS cross section is given by the distribution of partons carrying momentum fraction $x_B=Q^2/2m_p\nu$. The parton distributions can be expressed in terms of matrix elements of the nucleon target, \eg, for the quark distribution
\beqa \label{qdistr}
f_{\qu/N}(x_B,Q^2)&=& \frac{1}{8\pi} \int dy^- \exp(-ix_B p^+ y^-/2)
\label{melm}\\
&\times&\langle N(p)| \qb(y^-) \gamma^+\, {\rm P}\exp\left[ig\int_0^{y^-}
dw^- A^+(w^-) \right] \qu(0)|N(p)\rangle \nonumber
\eeqa
Here $y^- = y^0-y^3$ measures the `Ioffe' distance between the photon interaction vertices in the production amplitude and its complex conjugate. The Ioffe distance is along the light-cone (LC) since $y^+ = \morder{1/\nu}$ and $y_\perp =\morder{1/Q}$ vanish in the Bjorken limit. The path ordered exponential ensures the gauge invariance of the matrix element and physically represents final state Coulomb scattering of the outgoing (struck) quark. This scattering affects the DIS cross section, and hence the parton distributions, within the coherence length $\nu/Q^2 \sim y^-$ of the virtual photon. Note that it is essential for this picture that the virtual photon is moving (mainly) in the {\em negative} $z$-direction, $q^- \simeq 2\nu \sim 1/y^+ \to\infty$ in the Bjorken limit, whereas $q^+ = Q^2/q^- \sim 1/y^-$ is fixed.

Until recently the path ordered exponential was regarded as a gauge artifact that does not affect the cross section and can be eliminated by choosing the LC gauge $A^+=0$. I return to this question in the next section.

Integrating Eq. \eq{qdistr} over the momentum fraction $x_B$ and observing that the quark distribution is non-vanishing only for $-1\leq x_B \leq 1$ one derives sum rules which relate a {\em local} $(y^-=0)$ matrix element to moments of the quark distribution. Thus in the case of a nucleon with polarization $s$ one finds a sum rule for the quark helicity distribution
\beq \label{qspin}
s^\mu \int_0^1 dx \left[f_{\qu/N}^\uparrow(x_B,Q^2) - f_{\qu/N}^\downarrow(x_B,Q^2) + (\qu \to \qb) \right]
= \langle N(p,s)| \qb(0) \gamma^\mu \gamma_5 \qu(0)|N(p,s)\rangle 
\eeq
Measurements of DIS on longitudinally polarized targets, pioneered by the EMC collaboration \cite{emc}, have given the surprising result that quark helicities account for only a small fraction $\Delta\Sigma \simeq .23$ of the proton spin. The remaining spin must arise from quark orbital motion and gluons. Efforts to formulate a measurable DIS sum rule for the total angular momentum $J$ of the target have, however, not been successful (see \cite{disspin} for recent theoretical and experimental reviews).

\subsection{Deeply Virtual Exclusive Scattering}

The QCD factorization theorem has been extended to the case of exclusive final states, \eg, Deeply Virtual Compton Scattering (DVCS), $ep \to e\gamma p$ \cite{dvcsfact}. In the Bjorken limit $Q^2, \nu \to \infty$ at fixed $x_B$ the amplitude is dominated by the `handbag' configuration familiar from DIS, where the lower (target) vertex is now referred to as the Generalized Parton Distribution (GPD) \cite{gpv}. The DVCS handbag differs from the DIS one in several important respects:
\begin{itemize}
\item It is the amplitude for the specific $e\gamma p$ final state -- thus its absolute square gives the exclusive cross section.

\item The final photon is real, which requires that a total `plus' momentum $x_B m_p$ is transferred from the target. This momentum transfer is shared by the two partons connecting the photons to the target, leaving an integral over the fraction $x$ of plus momentum carried by the first parton. The struck quark line between the photon vertices is far off-shell except when the first parton transfers the full amount, $x=x_B$.

\item The GPD's are thus only indirectly given by the measured exclusive cross section, symbolically
\beq \label{sigpd}
\sigma(DVCS) \sim \left| \int_{-1}^1 dx\ GPD \right|^2
\eeq
In practice, it is necessary to construct models for the GPD's and then test how well they fit the data using the relation \eq{sigpd}.

\item The initial and final protons may have different transverse momentum and helicity. This implies the existence of two new GPD's, which appear multiplied with the momentum transfer in the handbag amplitude and which thus do not contribute to the forward amplitude measured in DIS.
\end{itemize}

 All GPD's are given by matrix elements analogous to Eq. \eq{qdistr}, where the initial and final nucleons have different momenta and spin (more generally, the final baryon may be an excited $N^*$). Hence the GPD's provide much new information on nucleon properties. 

The GPD's satisfy sum rules of the form \cite{gpv}
\beq \label{formsum}
\int_{-1}^1 dx\ GPD(x,x_B,t) = F(t)
\eeq
where $t$ is the invariant momentum transfer between the nucleons and $F(t)$ is the nucleon electromagnetic form factor. Moreover,
\beq \label{spinsum}
\int_{-1}^1 dx\ x GPD(x,x_B,t=0) = 2J
\eeq
where $J$ is the total spin of the nucleon and the GPD involves the new distributions which do not contribute to DIS. Hence the generalization to non-forward matrix elements allows a sum rule for the total angular momentum. Note that $t=0$ is outside the physical region, requiring an extrapolation of the data. The practical significance of Eq. \eq{spinsum} will depend on the reliability of the GPD models (both quark and gluon GPD's contribute to the proton spin). Some interesting phenomenology is already presented in section 4.2.3 of Ref. \cite{gpv}.

Models for the GPD's are constrained by the fact that they should reduce to the standard DIS distributions in the forward limit, and by sum rules like \eq{formsum}. First model predictions for single spin asymmetries \cite{kpv} are in qualitatively agreement with DVCS data from HERMES \cite{Hssa} and CLAS \cite{Cssa}.

I refer to the comprehensive review \cite{gpv} and these proceedings for more detailed discussions of deeply virtual exclusive processes and their description in terms of GPD's. These studies are very promising and experimentally challenging, motivating dedicated facilities. They have invigorated the study of hadron structure and remind us that there is much to learn about the proton besides its inclusive structure function. 

\section{Second step: DYNAMICS OF MATRIX ELEMENTS}

Given that we have measured hadron matrix elements -- preferably in several different processes to test factorization -- we need to address what these elements teach us about hadron structure. This question is somewhat vague, as the matrix elements represent long distance physics for which we lack a clear formalism in QCD. However, we may try to approach it in the spirit of the factorization theorems: Use perturbation theory as a model of soft dynamics, and take properties that hold at all orders as generally valid. Hadrons are then described as a superposition of quark and gluon Fock states, and expressions can be found for the matrix elements in terms of the Fock state wave functions.

I shall here briefly describe two issues related to the interpretation of matrix elements in which I have recently been involved.

\subsection{Parton Distributions are Not Probabilities}

There has been a general impression that the parton distributions measured in DIS can be simply interpreted as probabilities for finding a parton in the hadron. Thus interactions occurring after the virtual photon is absorbed on a target parton would not affect the DIS cross section. Recent work \cite{bhmps} shows that the situation is not that simple.

The argument that parton distributions are probabilities is deceptively simple. One chooses the light-cone (LC) gauge $A^+=0$ in the expression \eq{qdistr} for the quark distribution to reduce it to
\beqa  \label{qprob}
{\cal P}_{\qu/N}(x_B,Q^2) &=& \frac{1}{8\pi} \int dy^- \exp(-ix_B p^+ y^-/2) \langle N(p)| \qb(y^-) \gamma^+\,\qu(0)|N(p)\rangle_{y^+=y_\perp=0}\nonumber
\\&&\\
&=& \sum_n \int^{k_{i\perp}^2<Q^2}\left[ \prod_i\, dx_i\, d^2k_{\perp
i}\right] |\psi_n(x_i,k_{\perp i})|^2 \sum_{j=q} \delta(x_B-x_j)\nonumber
\eeqa
The second equality follows by noting that all states and fields are evaluated at the same LC time $y^+=0$. Thus when the nucleon state $|N,y^+=0\rangle$ is expanded on its quark and gluon Fock states $n$ the matrix element reduces to a sum of probabilities for states that contain a quark with momentum fraction $x_B$. See Ref. \cite{spd} for a detailed derivation in the case of Generalized Parton Distributions.

This argument contains a subtle but important flaw, which implies that the probability ${\cal P}_{\qu/N}$ of Eq. \eq{qprob} does not equal the parton distribution $f_{\qu/N}$ of Eq. \eq{qdistr}. $f_{\qu/N}$ is the leading twist result for the DIS cross section in the Bjorken limit for a general $(A^+\neq 0)$ gauge. The path ordered exponential in \eq{qdistr} arises from keeping only the asymptotically large terms $\propto q^- A^+$ in the rescattering of the struck quark on the target gluon field. This is correct in all gauges {\em except} LC gauge where this term vanishes. Thus the two limits $q^- = 2\nu \to \infty$ and $A^+\to 0$ do not commute. This occurs even in the Born term of elastic quark scattering (see Section 7 of Ref. \cite{bhmps}). The correct result is obtained by choosing LC gauge {\em before} the high energy limit.

It turns out \cite{jy} that there is a non-vanishing contribution to the path ordered exponential from asymptotic times in LC gauge. Using a specific prescription for the singularity of the LC gluon propagator $d_{LC}^{\mu\nu}(k)$ at $k^+=0$ this asymptotic contribution can be chosen to vanish. The final state interaction effects may then be (partly) mapped into a complex target wave function.

One may intuitively understand why the probability expression \eq{qprob} cannot contain all the physics of DIS scattering. The struck quark rescatters on its way out of the target, and the rescattering amplitudes must be added coherently within the coherence length $\nu/Q^2 = 1/(2m_p x_B)$ of the virtual photon. The contribution of on-shell intermediate states between the rescatterings gives the amplitudes complex relative phases. These phases are crucial for understanding diffraction and shadowing phenomena in DIS \cite{gg}. In the absence of rescattering contributions the amplitude is given by the (stable) target wave function, which contains no on-shell intermediate states and hence does not have dynamical phases\footnote{This holds in covariant gauges, and in LC gauge when the propagator singularity at $k^+=0$ is regularized using a principal value prescription.}. These arguments were explicitly verified by means of a perturbative model calculation in Ref. \cite{bhmps}.

The rescattering phase was subsequently shown \cite{bhs} to give a leading twist single spin asymmetry in semi-inclusive DIS,\ie, a correlation between the jet angle and the transverse polarization of the nucleon target in $e+N_\uparrow \to e + jet + X$. This `Sivers Effect' \cite{siv} was welcome in view of the large asymmetry observed experimentally \cite{Hssapi}. Such a correlation had been thought to be of higher twist, but the proof turned out to be incorrect since it ignored the complex phase arising from the path ordered exponential \cite{jcc}.

The universality of the rescattering corrections in different processes is presently under study. The single spin asymmetry in the Drell-Yan process should be reversed compared to DIS, since the phase arising from initial state rescattering is the complex conjugate of that from final state interactions \cite{jcc}. Very recently, the model used for studying shadowing in DIS was applied to the Drell-Yan process \cite{sp}. The result indicates that there could be essential differences between the rescattering corrections in the two processes.

While the dynamics of parton distributions thus appears considerably richer than previously anticipated, it should be noted that the rescattering effects discussed above are nevertheless fully determined by the target wave function. The struck quark does not have time to emit or absorb real, transverse gluons in the target. Consequently the quark does not influence the target gluon field, it only `samples' this field via Coulomb scattering. In a theory with only scalar interactions the rescattering effect would be absent, and the DIS cross section would be given by the parton probabilities analogous to Eq. \eq{qprob}.

From a quantitative point of view, the effects of rescattering on the cross section should be important mainly at low $x_B$ where the coherence length of the virtual photon is long. In particular, diffractive DIS and the shadowing of nuclear parton distributions are primary manifestations of rescattering.

\subsection{(Semi-) Exclusive Processes at Large t}

A perturbative QCD (PQCD) formalism for exclusive processes at large momentum transfer $|t|$, such as $ep \to ep$ (electromagnetic form factors) and fixed angle hadron scattering, was developed more than two decades ago \cite{lb}. The scattering amplitude factorizes into a hard (perturbative) scattering kernel and a `distribution amplitude' for each participating hadron. The distribution amplitude is given by the wave function of the lowest (valence quark) Fock state with a transverse size of order $1/\sqrt{|t|}$. Intuitively, only compact Fock states can absorb large momentum transfer without radiating particles.

Quantitative estimates of the scattering kernel have shown that it tends to be dominated by `endpoint' configurations where one valence quark carries most of the hadron momentum \cite{ilrk}. In this `Feynman mechanism' the fast quark absorbs all the momentum transfer, while the wee quarks have no preferential direction of motion and thus can form a bound state with the fast quark in both the initial and final states. The endpoint configurations are not compact, and should asymptotically be suppressed by Sudakov form factors. Their magnitude is sensitive to the shape of the distribution amplitude near its endpoints.

Data on exclusive cross sections obey `dimensional scaling' rules already at modest momentum transfers \cite{exscal}. This is in accordance with the hard QCD formalism \cite{lb}, even though the precocity of the scaling is surprising. The early scaling appears accidental in models based on endpoint contributions \cite{ilrk}, which are assumed to be asymptotically non-leading due to the Sudakov effects.

Polarization data is a sensitive indicator of the hardness of the reaction mechanism, since quark helicity is conserved in PQCD when the momentum transfer is large compared to the (effective) quark mass. In the exclusive PQCD formalism \cite{lb} the hadron helicity is given by the sum of the helicities of its valence quarks, implying hadron helicity conservation in hard processes.

Little polarization data has so far been available at large $t$. However, Jlab data on polarized $ep \to ep$ \cite{f2f1} suggests that the ratio of the proton flip to non-flip form factors $F_2(t)/F_1(t)$ decreases more slowly with $t$ than $1/(-t)$, which is the prediction of the exclusive QCD formalism \cite{lb}.

ZEUS recently published \cite{Zsemiex} striking data on polarization effects in the process $\gamma + p \to V + Y$, at large momentum transfer $|t| \lsim 12\ \gev^2$ between the projectile photon and the vector meson $V=\rho^0, \phi, J/\psi$. The mass of the target dissociative system $Y$ is small compared to the hadronic invariant mass $80 \leq W/{\rm GeV} \leq 120$. In this semi-exclusive kinematics the cross section is expected \cite{bdhp} to be governed by the hard subprocess $\gamma g + \to \qq + g$, with the quark pair forming the vector meson via its distribution amplitude. The dependence on the target proton is given by the inclusive gluon distribution evaluated at $x=-t/(M_Y^2-t)$.

The $\gamma + p \to \rho^0 + Y$ cross section (integrated over $x\geq 0.01$) scales as $d\sigma/dt \propto (-t)^n$, with $n = -3.21\pm 0.04 \pm 0.15$ in close agreement with $n= -3$ expected from dimensional counting. The corresponding data on $\phi$ production gives a power $n=-2.7\pm 0.1 \pm 0.2$. The ratio $\sigma(\phi)/\sigma(\rho) \simeq 2/9$ for $-t \gsim 3\ \gev^2$ in accordance with flavor SU$_3$. These features suggest that the subprocess is hard and described by PQCD.

Helicity conservation at the level of the hard subprocess requires that the quark and antiquark are created with opposite helicities. Hence the $\rho$ meson they form is predicted to be longitudinally polarized, $\lambda_\rho =0$. At the level of the external particles, on the other hand, helicity conservation implies that the $\rho$ meson have the same helicity as the (real) projectile photon, \ie, $\lambda_\rho = \pm 1$. We thus have a situation where helicity conservation at the external particle level is in conflict with helicity conservation at the quark level.

The ZEUS data \cite{Zsemiex} shows that $s$-channel helicity is nearly conserved in the entire measured range ($-t \leq 6\ \gev^2$) for both $\rho$ and $\phi$ mesons.  Hence helicity is violated at the subprocess level. In PQCD this brings a suppression factor proportional to the quark mass squared, $m_q^2/(-t)$. The cross section is then expected to scale with a power $n=-4$, whereas the data is closer to the dimensional counting rule $n=3$. Taken at face value, this means that the subprocess is soft (allowing helicity flip) yet obeys the dimensional counting rule. Given the other examples of precocious scaling the possibility that soft physics obeys dimensional counting merits consideration.

It is possible to directly measure the size of the subprocess from the dependence on the virtuality $Q^2$ of the projectile photon. For subprocesses with transverse size $\lsim 1/Q$ little dependence on the photon virtuality is expected. While there is as yet no data on the $Q^2$-dependence at large $|t|$, we have studied this in PQCD for the (somewhat simpler) subprocess $\gamma u \to \pi^+ d$ \cite{pltv}. The result is interesting. The amplitude for this process involves an integral over the momentum fraction $z$ carried by the $u$-quark of the form $\int dz\, \phi_\pi(z)/z$. Since the distribution amplitude vanishes at the endpoints, $\phi_\pi(z) \propto z$ for $z \to 0$, the amplitude is dominated by finite $z$ and the process appears compact. On the other hand, the derivative of the amplitude wrt. the photon virtuality $Q^2$ brings an additional factor of $1/z$ to the integrand, making the integral (logarithmically) singular at the endpoints. Hence $d\sigma/dQ^2dt$ is (formally) infinite at $Q=0$, corresponding to an infinite transverse size! Clearly the endpoint contributions are delicate, and merit further study.

\section{Third step: SOFT QCD}

In parallel with developing our understanding of hard QCD processes we need to consider also the soft, long-distance regime. The challenge is to devise a systematic approximation scheme for QCD which is both formally exact and phenomenologically relevant at low orders. The QCD origin of successful phenomenological descriptions such as the quark and Regge models needs to be demonstrated. 

Perturbation theory merits a particularly close look, given the similarity of the hadron spectrum with positronium and the indications that the soft and hard regimes are related (\cf\ Bloom-Gilman duality \cite{bg} and the dimensional scaling discussed above). The perturbative expansion is formally exact and very successful at short distances. In fact, it is the success of PQCD that makes us confident about the correctness of QCD.

Standard PQCD fails to describe prominent features of long distance physics, including chiral symmetry breaking and confinement. This is plausibly due to the expansion being based on an empty, perturbative vacuum. Although vacuum fluctuations are systematically included in the PQCD expansion, the effects of a gluon condensate cannot be mimicked at finite orders. 

However, we may choose to perturb around $|in\rangle$ and $\langle out|$ states that contain free quark and gluon pairs. Formally, any state at asymptotic time $t=\pm\infty$ relaxes to the true ground state at finite $t$ (assuming that there is a non-zero overlap). The partons in the asymptotic state only modify the Feynman $\ieps$ prescription of the  standard PQCD rules. A promising modification of the on-shell gluon and (massless) quark propagators was recently proposed \cite{crh},
\beqa
D_g^{ab,\mu\nu}(p) &=& -g^{\mu\nu} \delta^{ab}\left[\frac{i}{p^2+\ieps}+C_g
(2\pi)^4 \delta^4(p) \right] \label{gpropmod}\\
S_q^{AB}(p) &=& \delta^{AB}\left[\frac{i\Slash{p}}{p^2+\ieps}+C_q (2\pi)^4
\delta^4(p)\right]
\label{qpropmod}
\eeqa
which corresponds to adding gluons and massless quarks with vanishing 3-momenta to the asymptotic states. This prescription maintains Lorentz invariance at each order of the perturbative expansion. The PQCD expansion based on the modified propagators is formally as justified as the standard one, and agrees with it for short distance processes since the off-shell propagators are unchanged. 

The modified expansion captures more of the long-distance features of QCD at low orders in $\as$. Thus a finite gluon condensate $\ave{F_{\mu\nu}F^{\mu\nu}} \propto \as C_g^2$ arises through the 4-gluon coupling at two-loop level. The addition of $\qq$ pairs to the asymptotic states breaks chiral symmetry spontaneously, causing $\ave{\ol{\psi}\psi} \propto C_q$. At the same time, the vector and axial vector currents are conserved, ensuring gauge invariance and a massless pion. Further studies will show whether this approach can provide a real QCD description of hadron physics. To my mind the effects of modifying the boundary conditions of PQCD have not received the attention they deserve.

\end{document}